\documentclass[useAMS,usenatbib]{mn2e}
\usepackage{graphicx}
\def\aj{AJ}%
\def\araa{ARA\&A}%
\def\apj{ApJ}%
\def\apjl{ApJ}%
\def\apjs{ApJS}%
\def\ao{Appl.~Opt.}%
\def\aap{A\&A}%
\def\mnras{MNRAS}%
\def\qjras{QJRAS}%
\def\nat{Nature}%
\title[The dust temperatures of prestellar cores]
{The dust temperatures of the prestellar cores  in the $\rho$ Oph main cloud 
and in other star forming regions: consequences for the core mass function}
\author[D. Stamatellos, A.~P. Whitworth, and D. Ward-Thompson]
{Dimitris Stamatellos\thanks{E-mail:D.Stamatellos@astro.cf.ac.uk}, Anthony~P. Whitworth, 
and Derek Ward-Thompson\\ 
School of Physics \& Astronomy,Cardiff University, Cardiff, CF24 3AA, Wales, UK}

\begin{document}

\date{Accepted 1988 December 15. Received 1988 December 14; in original form 1988 October 11}

\pagerange{\pageref{firstpage}--\pageref{lastpage}} \pubyear{2007}

\maketitle

\label{firstpage}

\begin{abstract}
We estimate the dust temperatures of the clumps in the $\rho$ Oph main cloud 
taking into account the 3D geometry of the region, and external heating 
from the interstellar radiation field and from HD147879,
a nearby luminous B2V star, which is believed to dominate the radiation field
in the region.  We find that the regions where prestellar cores are observed
(i.e. at optical visual extinctions $>7$ mag) are colder than $\sim10-11$~K. 
These dust temperatures are
smaller than those which  previous studies of the same region have assumed.
We use the new dust temperatures to estimate the masses of the prestellar cores in the
 $\rho$ Oph main cloud from  mm observations, 
 and we find core masses that are larger than previous estimates
 by a factor of $\sim2-3$.  This affects the  core mass function (CMF) of the  
region; we find that the mass at which the core mass
spectrum  steepens from a slope $\alpha\sim1.5$ to a slope $\alpha\sim2.5$ has
moved from  $\sim0.5$~M$_{\sun}$ to  $\sim1$~M$_{\sun}$.
In contrast with the CMF in other star forming regions
(e.g. Orion), there is no indication
for a turnover down to the completeness limit ($\sim0.2$~M$_{\sun}$), 
but the CMF may flatten at around $\sim 0.4$~M$_{\sun}$.

We generalize our results to the prestellar cores in Taurus and in Orion. In
Taurus the ambient radiation field heating the prestellar cores is believed to be weaker 
than than that in $\rho$ Oph. Hence, the dust temperatures of the cores in Taurus
are expected to be below $\sim 10-11$~K.
In Orion the radiation field is believed to be $10^3$ times stronger than the
standard interstellar radiation field. Based on this assumption we estimate that the dust
temperatures of the prestellar cores in Orion are around $\sim 20-30$~K.

\end{abstract}

\begin{keywords}
Stars: formation -- ISM: clouds-structure-dust -- Methods: numerical -- Radiative transfer
\end{keywords}

\section{Introduction}  

Prestellar cores are condensations in molecular clouds that are on the verge 
of collapse or already collapsing (e.g. Myers \& Benson 1983;
Ward-Thompson et al. 1994, 1999; Ward-Thompson, Andr\'e \& Kirk 2002). 
They represent the first phase
in an evolutionary model of star formation that is based on observations of
different types of young objects: starless core/prestellar 
core
$\to$ Class 0 $\to$ Class I $\to$ Class II $\to$ Class III  
(Lada 1987; Andr\'e et al. 1993; Andr\'e et al. 2000; 
Di Francesco et al. 2007; Ward-Thompson et al. 2007).  
The study of starless/prestellar cores is important 
in constraining the initial conditions for star formation. 
Prestellar cores have been observed both in groups 
(as in e.g Ophiuchus, Taurus, Perseus) and in relative isolation 
(e.g. B68; Alves et al. 2001). They are very cold; their temperatures are below 20~K and
most probably around 10~K (Evans et al. 2001; Stamatellos \& Whitworth 2003a,b;
Stamatellos et al. 2004). They are also very dense; their typical
central densities are  $\stackrel{>}{_\sim}10^5 {\rm cm}^{-3}$. They
are observed either at NIR wavelengths
where they are seen in absorption against the luminous background
(e.g. Bacmann et al. 2000), 
or at FIR and submm wavelengths, where they appear in emission
(e.g. Motte et al. 1998; Ward-Thompson et al. 2002; Kirk et al. 2005). The peak
of their emission is around 150-200~$\micron$, consistent with the fact
that they are very cold.

Submm and mm observations are often used to determine the masses of cores
(e.g. Motte et al. 1998; Nutter et al. 2007). At these wavelengths the core
is optically thin to the radiation it emits, hence 
the observed flux  from the core is
$F_{\rm \lambda} =\tau_\lambda\ B_\lambda (T_{\rm dust})$. 
Therefore the column density $N(H_2)$ along the line of sight is
\begin{equation}
\label{eq:col.dens.gen}
N(H_2)=\frac{F_{\rm \lambda}}{\mu m_{H}\:\Delta\Omega\; \kappa_{\rm \lambda}\: 
B_{\rm \lambda} (T_{\rm dust})}\;,
\end{equation}
where 
$\Delta\Omega$ is the solid angle of the telescope beam for a resolved source,
or the solid angle of the source if unresolved,
$N(H_2)$ is the
hydrogen column density,
$\kappa_{\rm \lambda}$ is the dust opacity per unit mass, 
and $T_{\rm dust}$ the 
temperature of the dust (Hildebrand 1983). 
The above relation assumes that the dust is 
isothermal along the line of sight within the core. 
The mass of the core is then determined from the column density,
$
M_{\rm core}=\int{ N(H_2) dS}\,,
$
where the integral is over the projected area of the core. The main 
uncertainties in determining core masses using the above method come from 
our limited knowledge of the properties of the dust in cores 
($\kappa_{\rm \lambda}$) and the dust temperature ($T_{\rm dust}$). 
Eq.~\ref{eq:col.dens.gen} is very sensitive to the temperature,
since at these low temperatures the Planck function is non-linear
and the Rayleigh-Jeans approximation is not valid. Hence,
even underestimating or overestimating
the core temperature by a few degrees may lead one to 
overestimate or underestimate the Planck function (and consequently the core
mass) by a factor of 2 to 3 (Stamatellos \& Whitworth 2005b).

It is important then to investigate whether the core temperatures 
estimated by previous authors using the core SED (defined by
observations at only a few wavelengths) are actually representative of
the dust temperatures in cores. The aim of this paper is to use detailed
3D radiative transfer modelling, taking into account the core environment
(e.g. nearby luminous sources, ambient cloud), to estimate the 
temperatures in prestellar cores.
We will focus our study on the cores in the $\rho$ Oph 
main cloud, but we will attempt to generalise our results to other
star forming regions.

\section{The $\rho$~Oph main cloud}

$\rho$ Ophiuchi is a star forming region where many prestellar
cores have been observed (see Motte et al. 1998; Nutter et al. 2006)
along with more evolved protostars (Class 0, I, II objects; e.g.
 Motte et al. 1998; Bontemps et al. 2001; Wilking et al. 1989; Andr\'e
\& Montmerle 1994). It is 
one of the closest star-forming regions, being at a distance  
from 140 to 160~pc (Bontemps et al. 2001; Motte et al. 1998).

We  shall confine our study to the $\rho$ Oph main cloud
(i.e. L1688), where
six major clumps\footnote{To avoid confusion we note that
we use the term {\it clump} to refer to the larger
structures and the term {\it core} to refer 
to the condensations, i.e. the substructure, in these clumps.
Some authors (e.g. Motte et al. 1998) use the reverse terminology.}
 have been identified (Oph-A,
Oph-B, Oph-C ,Oph-D, Oph-E and Oph-F).
The prototypical Class 0 object VLA1623 
(Andr\'e et al. 1993) is located in the Oph-A clump.
Each of these clumps 
contains a few tens of solar masses and has an extent of $\sim 0.3$ pc.
These clumps show substructure; a large number 
($\stackrel{>}{_{\sim}}60$)  prestellar cores have
been identified in them. The core mass function
in the  $\rho$~Oph main cloud
is similar to the stellar IMF, which suggests that the masses 
of stars are determined by fragmentation at a very early stage
(Motte et al. 1998).

The external radiation field incident on these clumps is 
believed to be dominated by  HD147889, a nearby B2V star, which is located at a distance 
$0.5-1$ pc away from from the far side of the Oph-A clump (Liseau et al. 1999).
This is a luminous star of 5500 L$_\odot$ (Wilking et al. 1989). Assuming
a distance of 1 pc and a clump radius of 0.2 pc, if there is no
attenuation, 1\% of the star's radiation will heat the clump, i.e. 
55  L$_\odot$. This radiation dwarfs any radiation from other sources
(e.g. the interstellar radiation field, nearby young protostars).

Observations of the wider area of the main $\rho$ Oph cloud 
($\sim 4\ {\rm deg}^2$)
using SCUBA on the JCMT and comparison with 2MASS extinction
maps as part of the COMPLETE project (Ridge et al. 2006) reveal 
that there are no prestellar
cores (at least down to the SCUBA detection limit)
at visual extinctions $<7-10$ mag
(Johnstone et al. 2004). It is presumed that the 
UV radiation present in regions of low visual extinction makes
these regions hostile for star formation (McKee 1989). 
Surveys in other star forming regions have also revealed similar extinction
thresholds that appear to be connected to the environment of the region 
(e.g. in Perseus the extinction threshold is lower, $A_{\rm V}>5-7$ mag;
Kirk et al. 2006).

Thus, it appears that most cores are quite embedded in their parent
molecular clouds. 
Previous radiative transfer simulations of embedded cores
suggest that such cores are very cold 
(Stamatellos \& Whitworth 2003a,b; Stamatellos et al. 2004).
The temperatures at their 
centres is around 6~K and at their edges is from around 11~K (for a core
embedded in a parent cloud of $A_{\rm V}=5$) to 9~K 
(for a core embedded in a parent cloud of $A_{\rm V}=20$). This
is because the deeper the core is embedded the lower is
the strength of  the radiation field that heats the core.
Hence, it is appropriate to ask whether previous studies of 
prestellar cores have assumed core temperatures that are consistent with 
the fact that these cores are deeply embedded in their parental 
clouds.

We focus our study initially on  the $\rho$ Oph region.
Our goal is to construct a radiative transfer model to
estimate the temperature of the cores in this region
taking into account the 3D geometry of the region and 
the role of HD147879 in externally heating the clumps. 
We will discuss the applicability of 
this study to other regions of star formation
where low-mass stars form, and to 
more energetic regions of star formation, such as the
Orion nebula, by discussing the role of a stronger external 
radiation field. 

The structure of the paper is as follows. In Section~3
we discuss the radiative transfer method we use and
the constituents of the model (i.e. radiation sources, 
model geometry, dust properties, clump density profile). 
In Section~4 we describe the results of the radiative
transfer modelling with respect to the temperature
profile and the SEDs of the clumps in $\rho$ Oph, and in Section~5
we discuss the implication of the estimated dust temperature
for the core mass function of  $\rho$ Oph. In Section~6, we 
discuss the dust temperatures of cores in 
other star forming regions, and finally 
in Section~7 we summarize our results.

\section{The model}

\begin{figure}
\centerline{
\includegraphics[width=6.7cm]{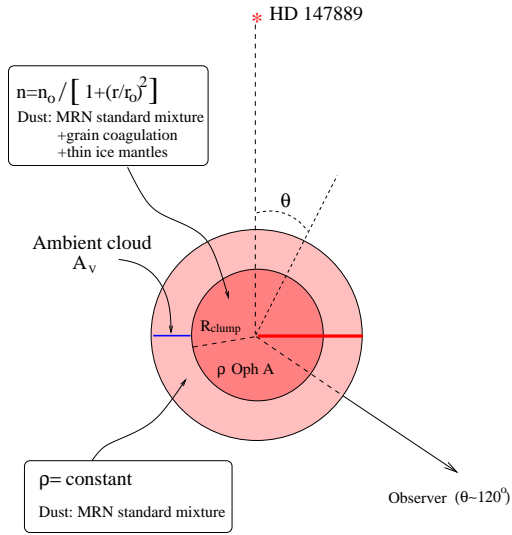}}
\caption{Schematic representation of the $\rho$~Oph~A model. 
The $\rho$ Oph A clump is surrounded by a virtual ambient cloud that
modulates the incident radiation that heats the clump.  The external heating
is provided both by the interstellar radiation field and by HD147889, a 
nearby B2V luminous star. The general geometry
of the models for the other $\rho$ Oph clumps are similar to this one.}
\label{fig:model}
\end{figure}

The big clumps of  $\rho$ Oph are represented by Plummer-like
density profiles, i.e. we invoke a spherical geometry, with 
the density being approximately flat in the centres of clumps and dropping as 
$r^{-2}$ in the envelopes. We exclude the Oph-B1 clump,
which is quite flattened, and the Oph-E clump, which appears to
be part of Oph-C.  We shall assume that  the cores in the clumps
are superimposed on this density profile, without actually taking
them individually into account in this model. However,
in Section~\ref{sec:sph} we shall estimate their effect on the dust temperatures.
We also assume that these clumps are heated (i) by the interstellar
radiation field and (ii) by radiation from the  HD14789 B2V star 
(see Fig.\ref{fig:model}). We
will adopt the 3D model constructed by Liseau et al. (1999).
Hence, in our model these clumps define the optical depth
that the external radiation has to penetrate to reach the individual cores
embedded inside the clumps.
In the next subsections we describe in detail the model.

\subsection{Density profiles}

Each of the $\rho$~Oph  clumps is represented by a density profile,
\begin{equation}
\label{plummer}
n(r) = n_0\,\frac{1}{ 1 + \left( \frac{r}{r_0} 
\right)^2 } \,,
\end{equation}
where $n_0$ is the density at the centre of the clump, and 
$r_0$ is the extent of the region in which the density is 
approximately uniform. 
This  profile is approximately flat in the central region and  falls of as $r^{-2}$
in the envelope, similarly to the more commonly used 
Bonnor-Ebert sphere density profile (Ebert 1955; Bonnor 1956). 
Additionally, despite being ad hoc,
it fits well the observed profiles of prestellar cores 
and it  predicts lifetimes, accretion rates, collapse velocity fields, 
SEDs and isophotal maps which agree well with observation, using
a minimum number of free parameters 
(Whitworth \& Ward-Thompson 2001). 
The adopted values for  $n_0$, $r_0$ and the extent $R_{\rm clump}$ of each clump
are listed in Table~\ref{tab:model.params}. These values are consistent with observations
of these clumps (e.g. Motte et al. 1998; Andr\'e et al. 1993).
We assume that the cores in the  clumps
are superimposed on this density profile (we re-iterate that at this stage we 
ignore the sub-structure of the clump). 

Each clump is surrounded by a virtual ambient cloud, which has a uniform density.
The role of these ambient clouds is to modify 
the ambient radiation field that heats each clump externally.
The optical extinction through the ambient cloud is chosen so that the peak of
computed SED of each clump corresponds to a dust temperature that matches
the dust temperature  of each clump assumed by previous authors.
The optical extinctions of the assumed ambient clouds around each one of
the $\rho$ Oph clumps are also listed in  Table~\ref{tab:model.params}. 

\subsection{Dust properties}

\begin{figure}
\centerline{
\includegraphics[width=7.5cm]{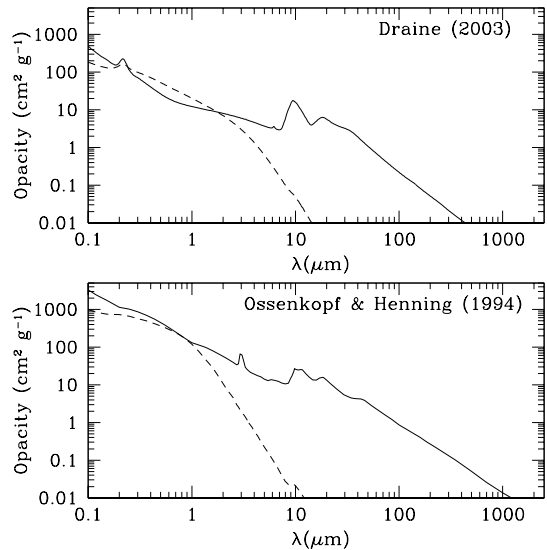}}
\caption{Dust opacity for the clumps (Ossenkopf \& Henning 1994) and
the virtual ambient cloud (Draine 2003). The solid lines correspond to
the opacity due to absorption, and the dashed lines to the opacity
due to scattering.}
\label{fig:opa}
\end{figure}

We use two kinds of dust; one for the clumps, and one for the ambient cloud
(Fig.~\ref{fig:opa}).
The dust grains in dense clumps are expected to coagulate and 
accrete ice mantles, so we use the Ossenkopf \& Henning (1994) opacities for 
MRN dust that has coagulated and accreted thin ice mantles for a period 
of $10^5$ years at a density $10^6$~cm$^{-3}$.
The ambient cloud has relatively low density and it can be considered 
as a part of the interstellar medium. Hence, for this layer a 
standard MRN dust opacity is used (Draine 2003). 

\begin{table*}
\begin{minipage}{0.9 \textwidth}
\caption{(I) Assumed parameters for each clump (based on observations). 
(II) Results of the radiative transfer modelling.}
\label{tab:model.params}
\centering
\renewcommand{\footnoterule}{}  
\begin{tabular}{@{}clccccc} \hline
\noalign{\smallskip}
& Parameter 		& $\rho$~Oph A &  $\rho$~Oph B-2 &  $\rho$~Oph C &  $\rho$~Oph D &  $\rho$~Oph F \\
\noalign{\smallskip}
\hline
\noalign{\smallskip}
&$R_{\rm clump}$
\footnote{Clump extent.}
 (AU)	& $2\times10^4$ & $2.9 \times10^4$	&$1.7\times 10^4$ 	&  $1.2\times 10^4$	& $1.2\times 10^4$ \\
 &$n_{\rm 0}$ 
 \footnote{Clump central number density.}
 (cm$^{-3}$)	& $16\times10^6$&  $1.2\times10^6$&$1.6\times10^6$	& $5.2\times10^5$ 	& $3.3\times10^5$ \\
  &$r_0$
  \footnote{Flattening radius of each clump.} 	
  (AU)		& $10^3$     	& $5.4\times 10^3$ 	& $3.4\times 10^3$ 	& $4\times 10^3$	& $7\times 10^3$ \\
\ (I)  &$A_V^{\rm cloud}$
\footnote{Visual extinction through the ambient cloud.}
	& 0.19 		&0.35  			& 2.8			& 0.45	 		& 0.17 \\
  &$d_\star$
  \footnote{Distance of each clump from HD147889}
   (pc)		& 1.1		&1.4 			& 1.3			& 1.8	 		& 1.6 \\
 &$M$
 \footnote{Clump mass (calculated from the density profile assumed and the extent of the clump).} 
 (M$_{\sun}$)	& 24 		& 61 			& 18 			& 47 			&  14\\
 &$A_V^{\rm clump}$
 \footnote{Visual extinction to the centre of each clump (calculated from the density profile assumed and the extent of the clump).}
 	& 1090 		&400  			& 560 			& 120			& 130\\ 
\noalign{\smallskip}
\hline
\noalign{\smallskip}
  &$T_{\rm centre}$
  \footnote{Temperature at the centre of the clump.}
   (K)		& 6 	& 7  & 7  & 7 & 8 \\
\ (II) &$T_{\rm edge}^{\rm min}$
\footnote{Temperature at the edge of the clump facing away from the H147989 direction.}
 (K)	& 14	& 13 & 12 & 14 & 14 \\
  &$T_{\rm edge}^{\rm max}$
 \footnote{Temperature at the edge of the clump facing towards H147989.}
   (K)	& 27 	& 22 & 14 & 20 & 24 \\
  &$\lambda_{\rm peak}$ 
 \footnote{Wavelength of the peak of the computed SED}($\micron$)
   	& 125	& 200 & 270 & 200 & 160 \\

\noalign{\smallskip}
\hline
\end{tabular}
\end{minipage}
\end{table*}

\subsection{Radiation sources \& geometry of the model}

Each clump is heated (i) by the interstellar radiation field, attenuated
through the ambient cloud, and (ii) by HD147889, a B2V nearby star which is located
at distance $\sim0.5 -1 $~pc from the $\rho$~Oph region (e.g. Liseau et al. 1999). 
In this model we will ignore radiation from deeply embedded protostars in these
clumps (e.g. VLA 1623, IRS43). According to the Stamatellos et al. (2005)
model the presence of an embedded protostar heats only the region
around it, hence the results presented here are not greatly affected by
the presence of young protostars in or around the clumps.
In Section~\ref{sec:protostars} we shall discuss the effects of such embedded protostars.

For the stellar radiation we adopt the parameters computed by  Liseau et al. (1999)
($T_\star^{\rm eff}=22,000$~K, $R_\star=5\ {\rm R}_{\sun}$, 
$L_\star=5,300\ {\rm L}_{\sun}$). We shall assume that this star emits as
a blackbody having temperature 22,000~K.
Due to the presence of this star in the vicinity of $\rho$~Oph, the external heating
of the clumps is highly anisotropic.

For the external interstellar radiation field  we adopt a revised version of the Black 
(1994) interstellar radiation field (BISRF). The BISRF consists 
of radiation from giant stars and dwarfs, thermal emission from dust grains, 
cosmic background radiation, and mid-infrared emission from transiently heated 
small PAH grains (Andr\'e et al. 2002). This radiation field  is modulated by
the ambient cloud around each clump, hence the incident radiation field on each of
the clumps is enhanced at FIR and longer wavelengths, and attenuated a shorter
wavelengths.

The geometry of the $\rho$~Oph region is taken from the 3D model
constructed by Liseau et al. (1999) based
on far infrared spectrophotometric observations with the ISO-LWS. Using
this model and the 1.3mm mosaic image of
Motte et al. (1998), we estimate the distance of HD147889 from each of the $\rho$~Oph
clumps (see Table~\ref{tab:model.params}).
These are rough estimates  but they do not greatly affect 
the results of the model.

\subsection{Monte Carlo radiative transfer}

The radiative transfer calculations are performed using {\sc PHAETHON}, 
a 3D Monte Carlo radiative transfer code developed by  Stamatellos \& 
Whitworth (2003). The code uses a large number of monochromatic luminosity
packets to represent the radiation sources in the system. The luminosity packets
are injected into the cloud and interact (are absorbed, reemitted, scattered) 
stochastically with it.  If an $L$-packet is absorbed its energy is 
added to the local region and raises the local temperature. To ensure 
radiative equilibrium  the $L$-packet is re-emitted immediately with 
a new frequency chosen from the difference between the local cell 
emissivity before and after the absorption of the packet (Bjorkman 
\& Wood 2001; Baes et al. 2005).

The model for each $\rho$ Oph clump is essentially 2-dimensional, 
hence the code used here is adapted and optimised for the study of systems 
with azimuthal symmetry. 
Each clump is divided into a number of cells by spherical  and conical surfaces. 
The spherical surfaces are evenly spaced in radius, and there are typically 100 of them. 
The conical surfaces are evenly spaced in polar angle, and there are typically 40 of them. 
Hence, each  clump is divided into $\sim4000$ cells. The number of cells used 
is chosen so that the density and temperature differences between 
adjacent cells are small.

The $L$-packets representing the ambient radiation field 
(typically a few $10^{10}$ packets)
are injected from the outside of the clump with injection  
points and injection directions chosen to mimic an isotropic radiation 
field incident on the ambient cloud around each clump (Stamatellos et al. 2004).

The $L$-packets representing the stellar radiation from  HD147889 
(typically a few $10^{10}$ packets are used) are emitted from the star with random 
direction and only a fraction of them ($\sim 1\%$) heats the clumps.
However, due to the large luminosity of  HD147889, this small percentage of
stellar photons dominates over the background radiation field.

\section{Dust temperatures and SEDs of the clumps in the $\rho$ Oph main cloud}
\label{newtemps}

The optical depth of the virtual ambient cloud around each clump is varied so that to
produce an SED for each clump that  peaks at a wavelength corresponding 
to a dust temperature that matches the dust temperature assumed by previous authors.
In the next subsections we
present the dust temperature profiles and the SEDs  of each clump as calculated 
by this model (see Figs.~\ref{fig:ophA}-\ref{fig:ophF}).

\subsection{$\rho$ Oph A}

\begin{figure} 
\centerline{
\includegraphics[width=10cm]{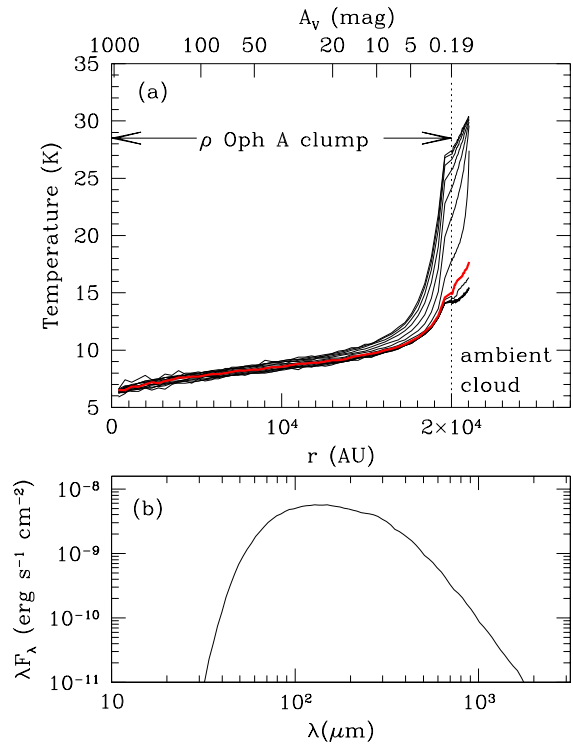}}
\caption{
{\bf (a)} Dust temperature profile of the Oph-A clump versus the distance from its 
centre, and versus visual extinction from the surface of the clump.
We plot the radial temperature
profile from  $\theta=0\degr$ to $180\degr$ to every $9\degr$, where
 $\theta=0\degr$  corresponds to the direction towards the B2V star.
The red thick line corresponds to the direction
perpendicular to the clump-star direction  (red thick strip in Fig.~\ref{fig:model}).
{\bf (b)} Simulated SED of the clump.
}
\label{fig:ophA}
\end{figure}

\begin{figure} 
\centerline{
\includegraphics[width=10cm]{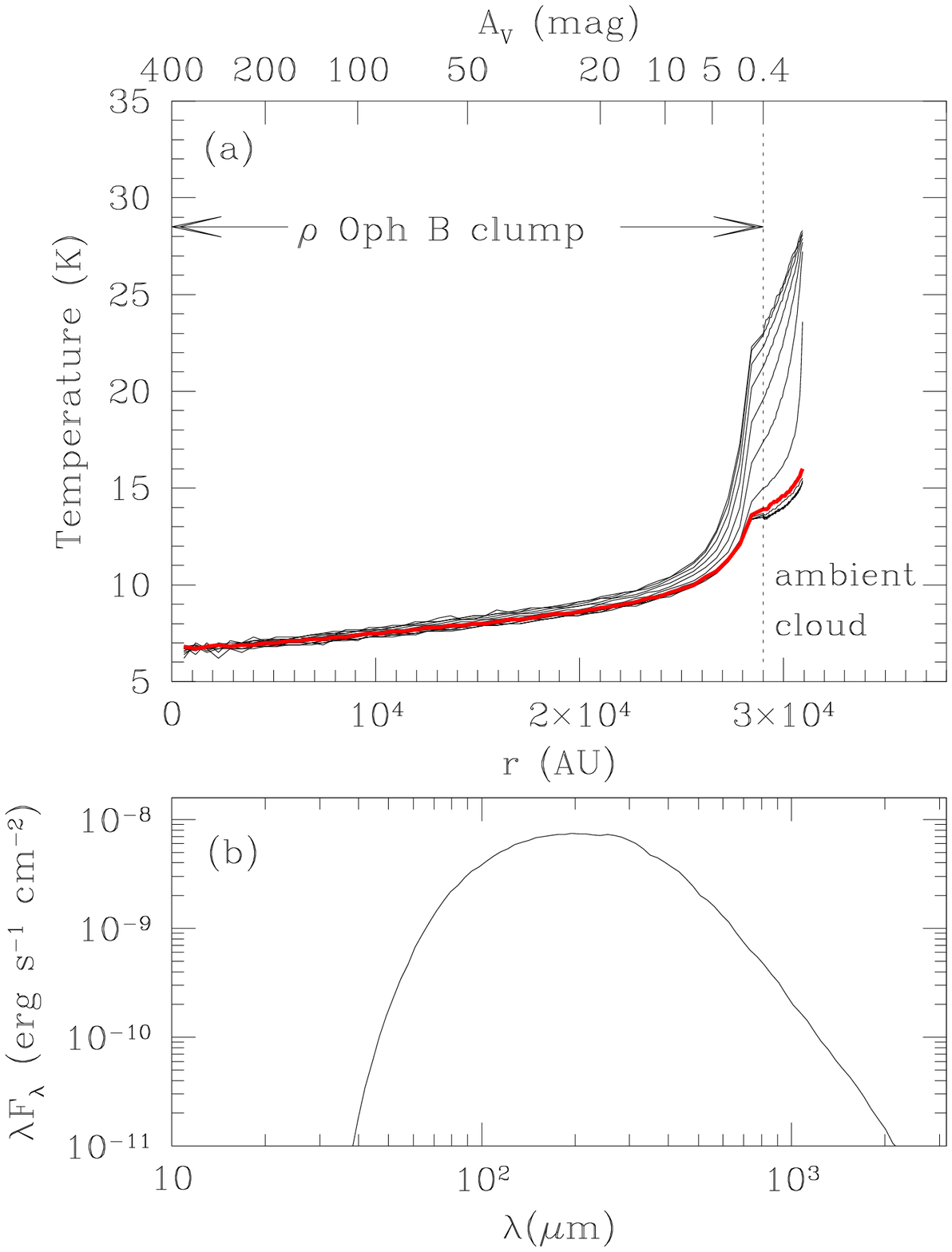}}
\caption{Same as in Fig.~\ref{fig:ophA} but for the Oph-B2 clump.}
\label{fig:ophB}
\centerline{
\includegraphics[width=10cm]{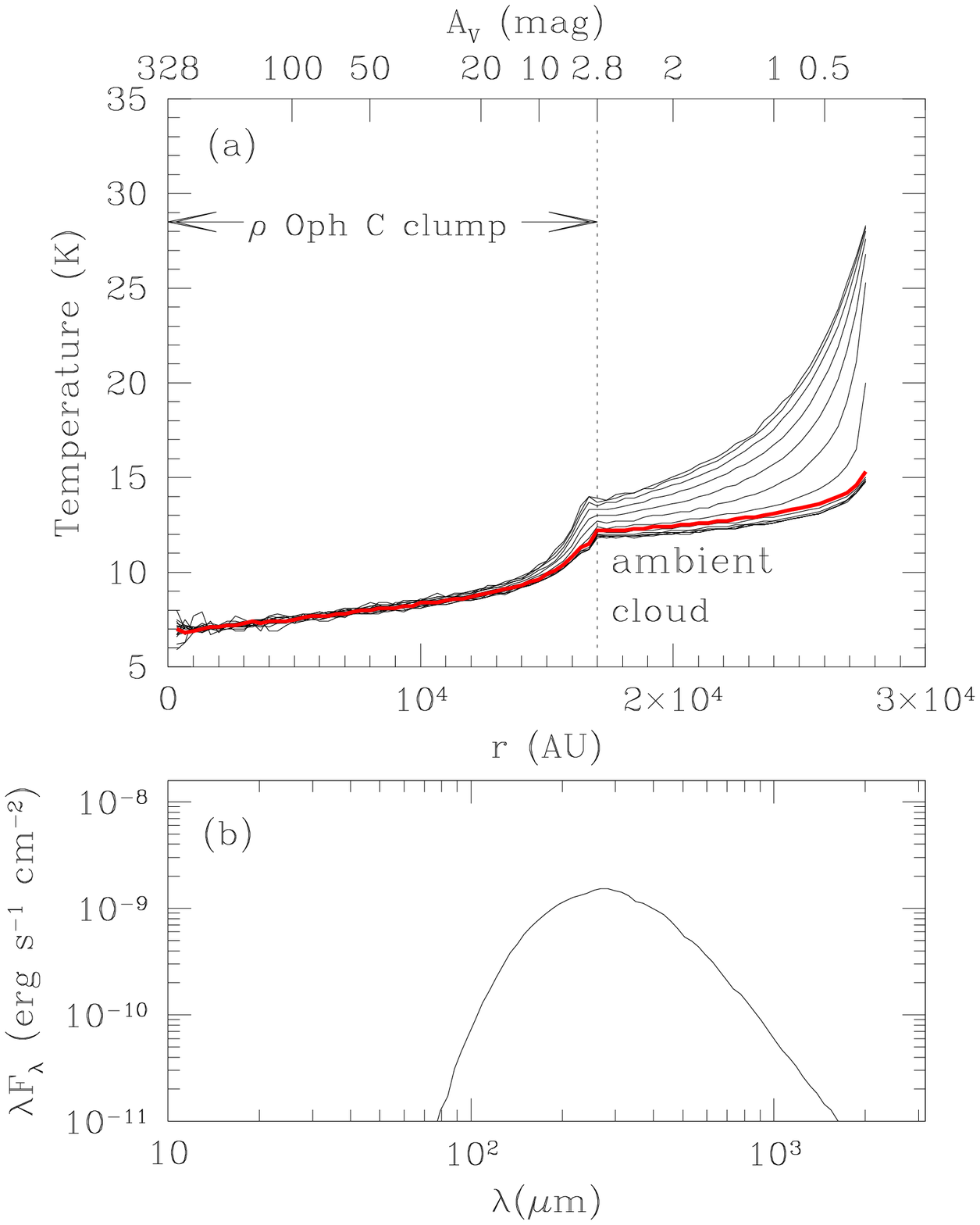}}
\caption{Same as in Fig.~\ref{fig:ophA} but for the Oph-C clump}
\label{fig:ophC}
\end{figure}

\begin{figure} 
\centerline{
\includegraphics[width=10cm]{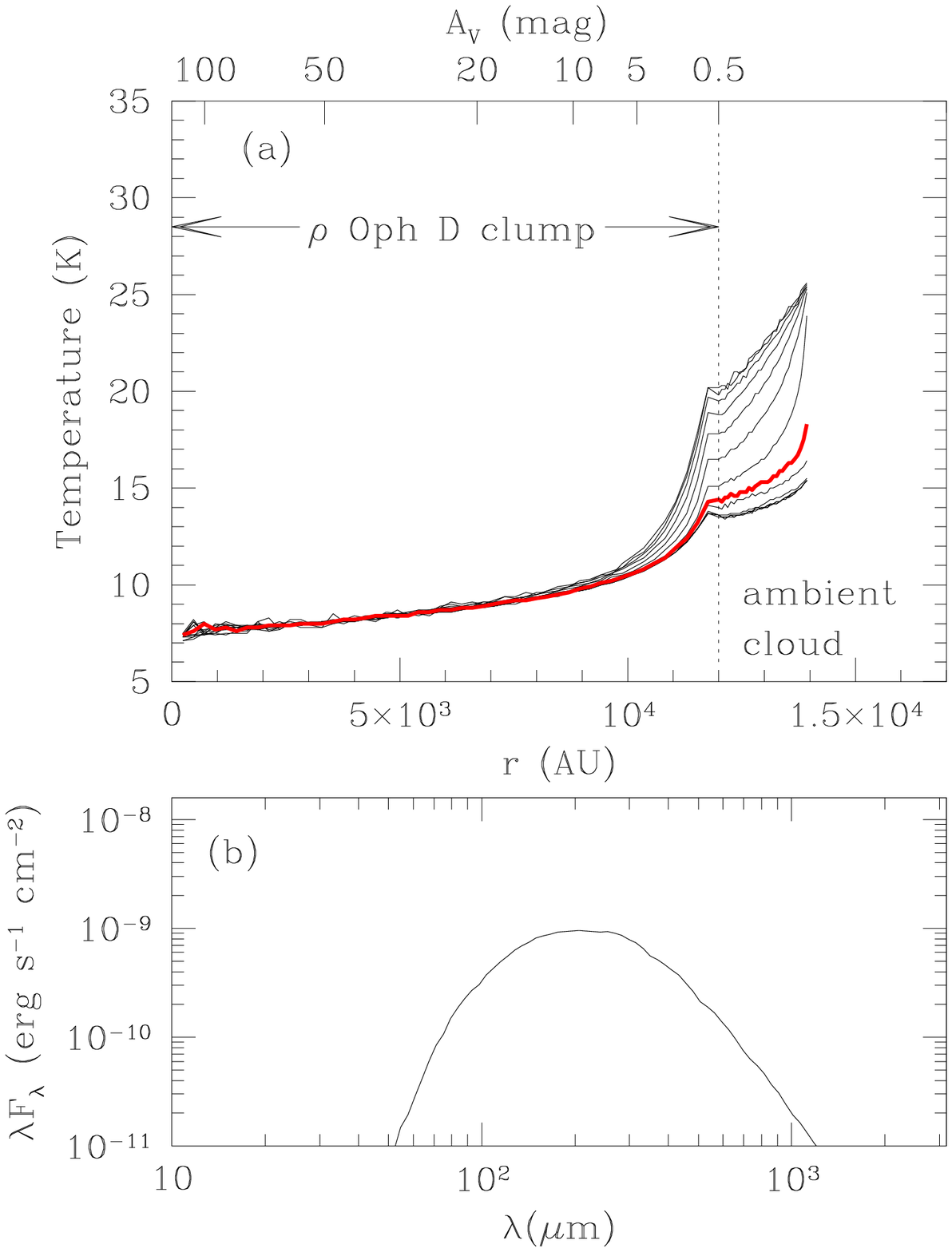}}
\caption{Same as in Fig.~\ref{fig:ophA} but for the Oph-D clump.}
\label{fig:ophD}
\centerline{
\includegraphics[width=10cm]{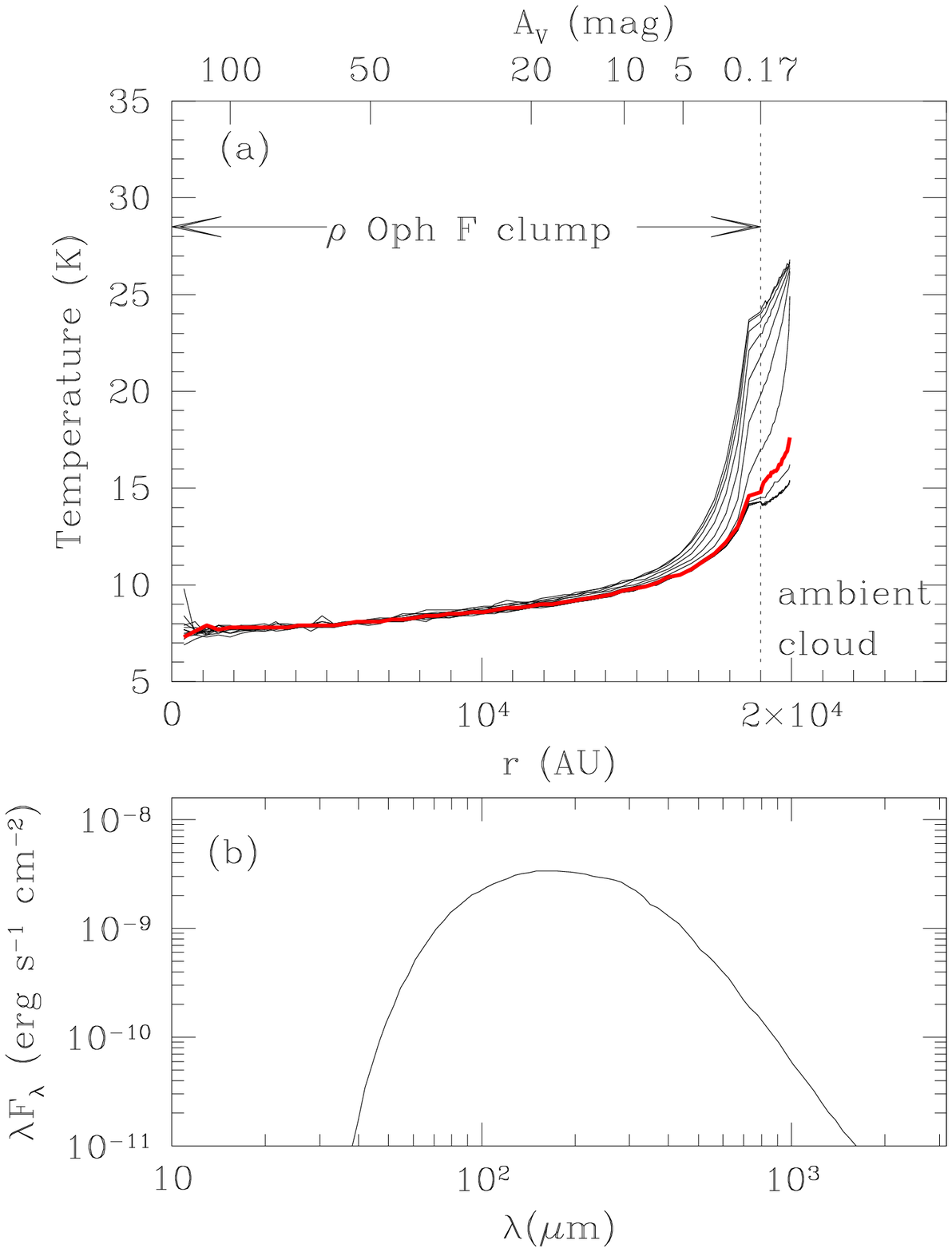}}
\caption{Same as in Fig.~\ref{fig:ophA} but for the Oph-F clump.}
\label{fig:ophF}
\end{figure}

Oph A is, according to the Liseau et al. (1999) 3D model, 
the closest clump to HD147889. 
The radiation from the star dominates over the interstellar radiation field 
and heats the clump to temperatures that range from $6-7$~K in the centre of
the clump, to $13-27$~K at the clump edge (Fig.~\ref{fig:ophA}). 
The clump hemisphere closer to the
star has higher temperatures (up to 27~K) than the other hemisphere, which has
temperature not higher than 15~K. In total, 80\% of the clump volume 
is colder than 15 K. The ambient cloud has dust temperatures 30-40~K 
on the side closer to the star, which are consistent with previous
temperature estimates based on the  PDR model  of Liseau et al. (1999). 

The SED of the model of $\rho$~Oph~A 
(Fig.~\ref{fig:ophA}b) peaks at around 125~$\micron$,
which is suggestive of a 
temperature\footnote{We note that the peak of SED corresponds to  the peak
of $\lambda\ k_\lambda B_\lambda$, which for  
 $k_\lambda\propto \lambda^{-1.78}$, i.e. the Ossenkopf \& Henning (1994) opacity,
translates to $\lambda_{\rm peak} T \approx 2,500\ {\rm K\ \micron}$.}
around 
$\sim 20$~K.
This value is consistent with what other studies have indicated 
(e.g. Andr\'e et al. 1993)
about the temperature derived from the SED. However, the peak of
the SED peaks at this wavelength due to the contribution from the outer hotter
parts of the clump, whilst most of the clump is colder.
The region of the clump where prestellar cores are observed, i.e. at visual extinctions
$A_{\rm V}>7$ mag (according to Johnstone et al. 2004), is colder than 11~K.

\subsection{$\rho $ Oph B-2}

The temperature profile of the $\rho $ Oph B-2 
clump (Fig.~\ref{fig:ophB}) is similar to that of Oph A. This clump
is colder than Oph A, and its 
temperature drops from $13-23$~K at the edge of the cloud to
$\sim7$~K at its centre. 
The SED distribution  of the clump peaks at around 195~$\micron$
indicating a temperature of 13~K, which is consistent with previous observations.
As in Oph A, the SED peak is characteristic of the outer warmer layers, whereas
most of the clump is colder than 13~K. Again assuming that cores 
are observed at visual extinctions $A_{\rm V}~>~7$~mag,
we find that their temperatures should be below 10~K.

\begin{figure}
\centerline{\includegraphics[width=4.9cm,angle=-90]{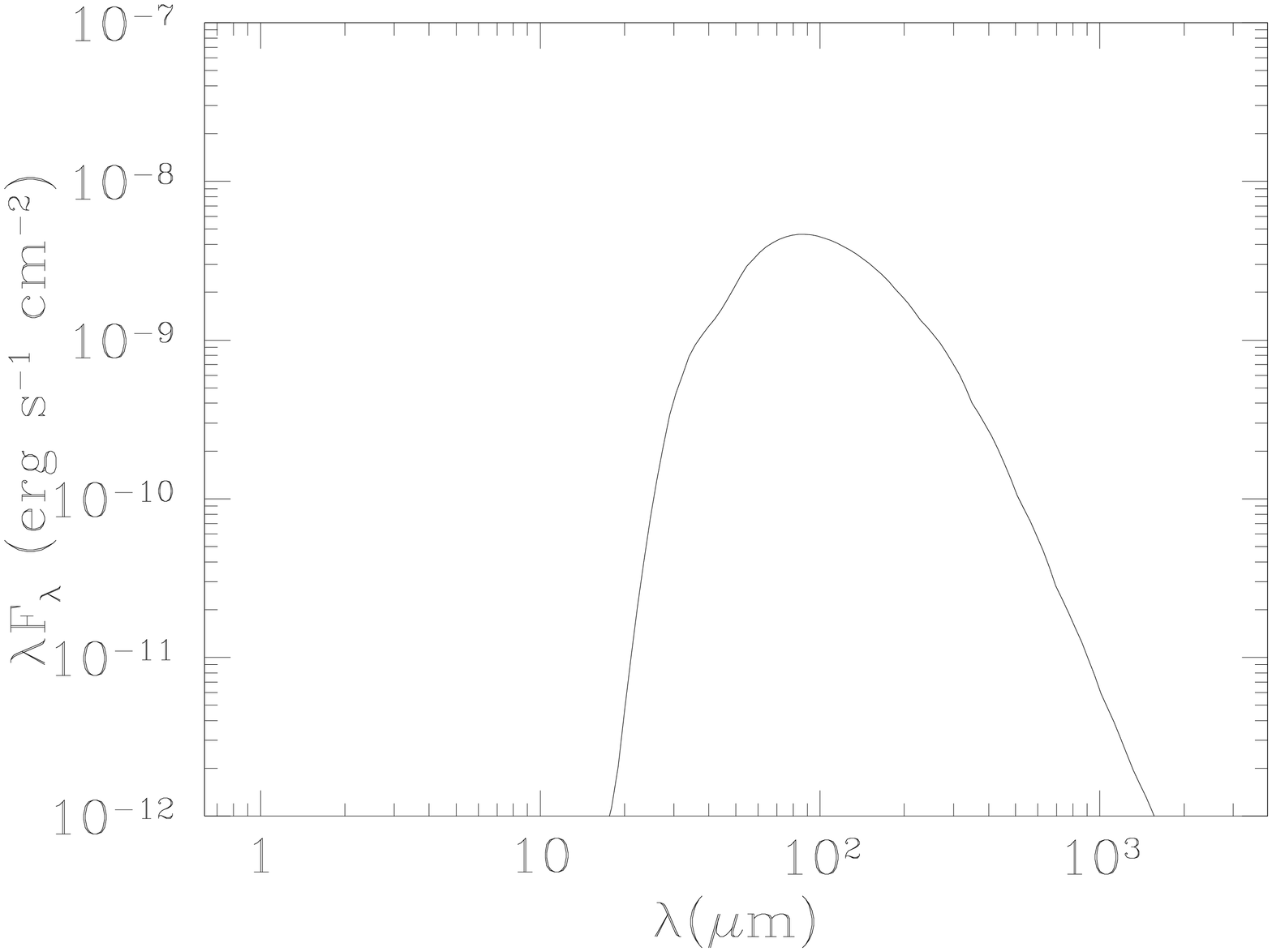}}
\caption{SED of a young Class 0 object as calculated by
Stamatellos et al. (2005), assuming a distance of 160~pc. The
bolometric luminosity of this source is 5~L$_{\odot}$.}
\label{fig:class0.sed}
\centerline{\includegraphics[width=7.cm]{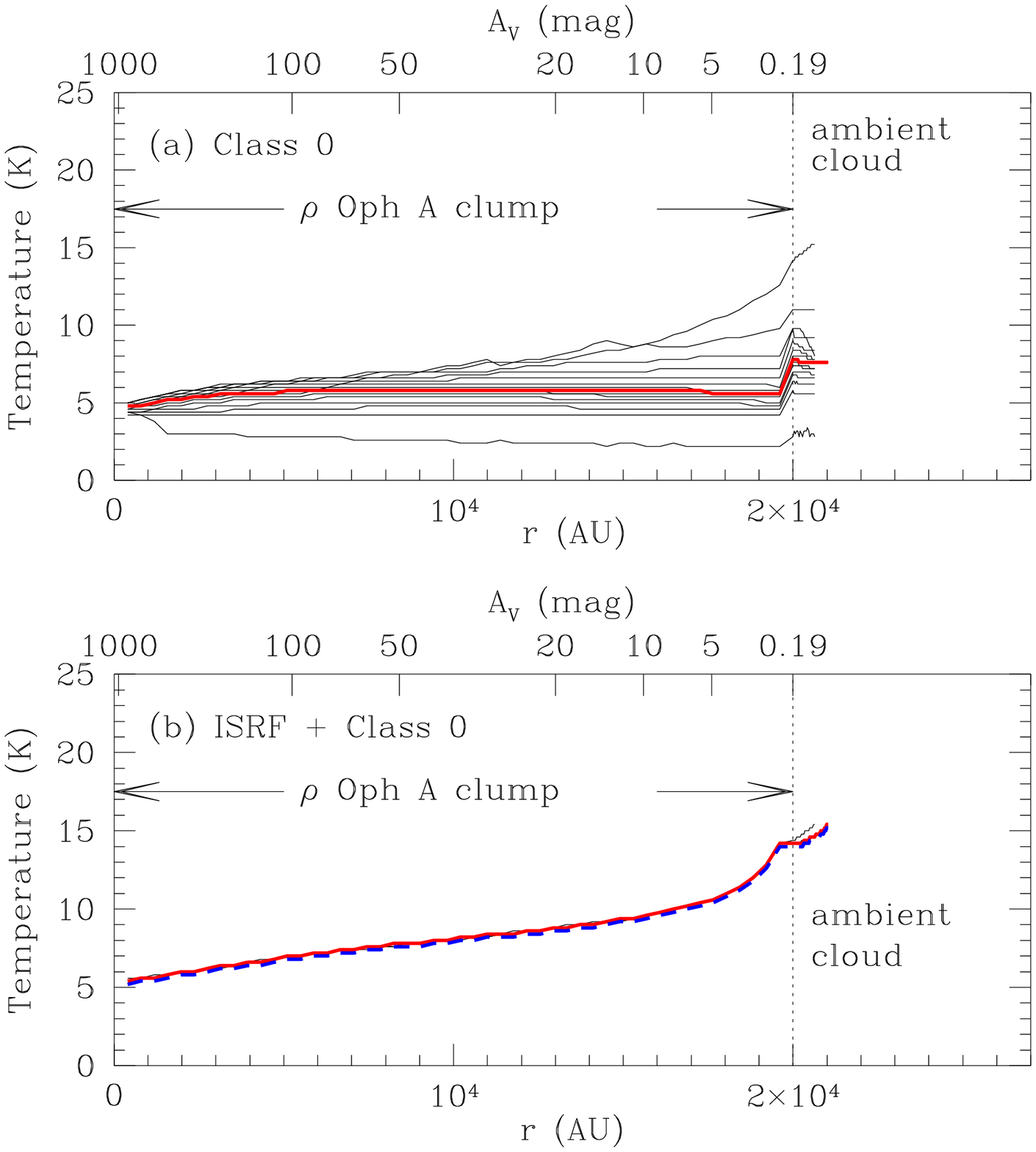}}
\caption{(a) Dust temperature profile of the clump, in different
directions, due to heating from a Class 0 object at the edge of the clump
(b) Dust temperature profile, in different
directions, due to heating from a Class 0 object and the interstellar
radiation field. The thick (red) line corresponds to the direction
perpendicular to the clump-protostar direction. The dashed (blue) line
corresponds to the temperature when only heating from the ISRF is considered.
The role ISRF is dominant. }
\label{fig:class0}
\end{figure}

\begin{figure}
\centerline{\includegraphics[width=4.7cm,angle=-90]{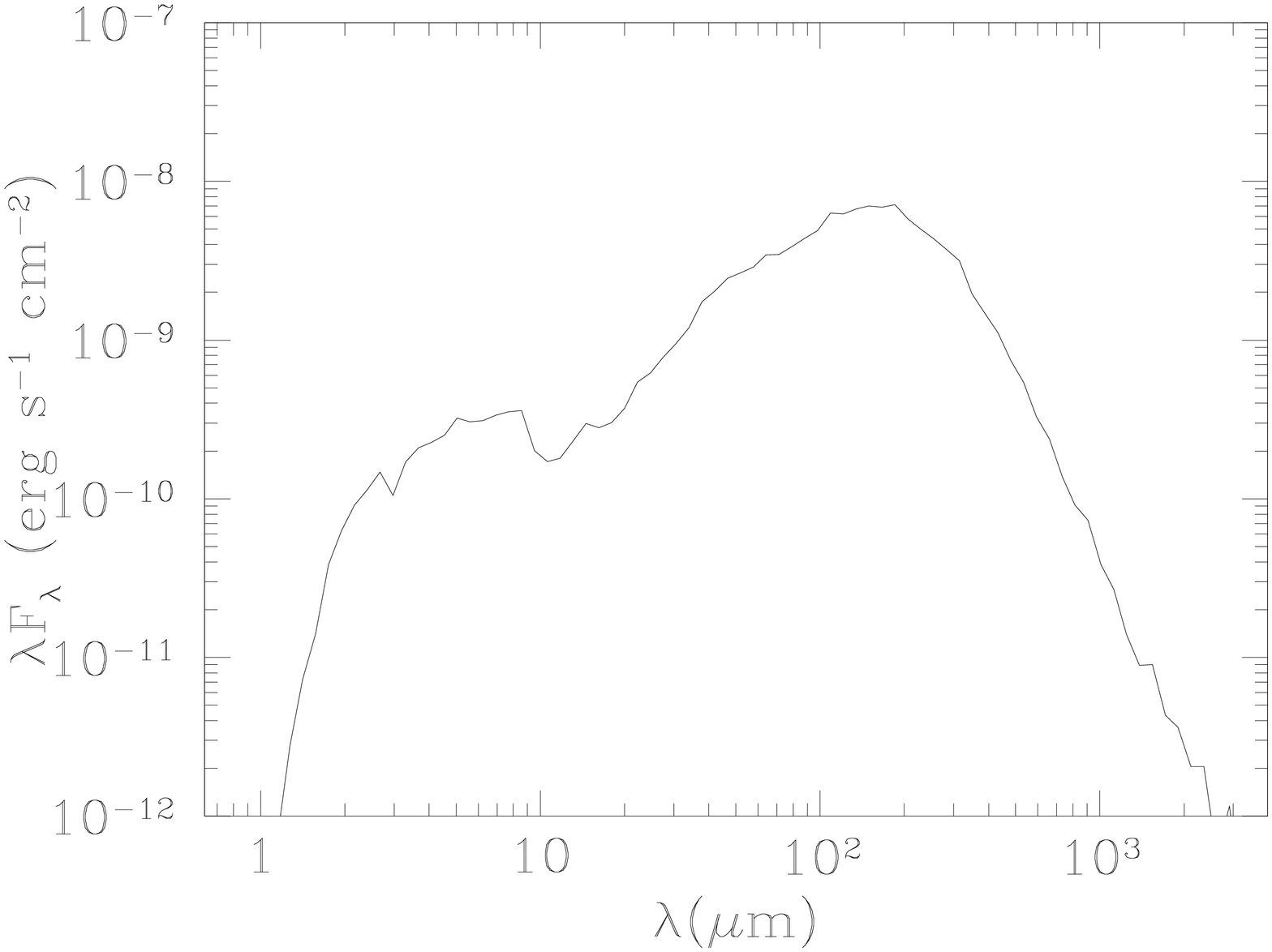}}
\caption{SED of a young Class I object as calculated by
Stamatellos et al. (2005), assuming a distance of 160~pc. The
bolometric luminosity of this source is 9.3~L$_{\odot}$.}
\label{fig:classI.sed}
\centerline{\includegraphics[width=7.cm]{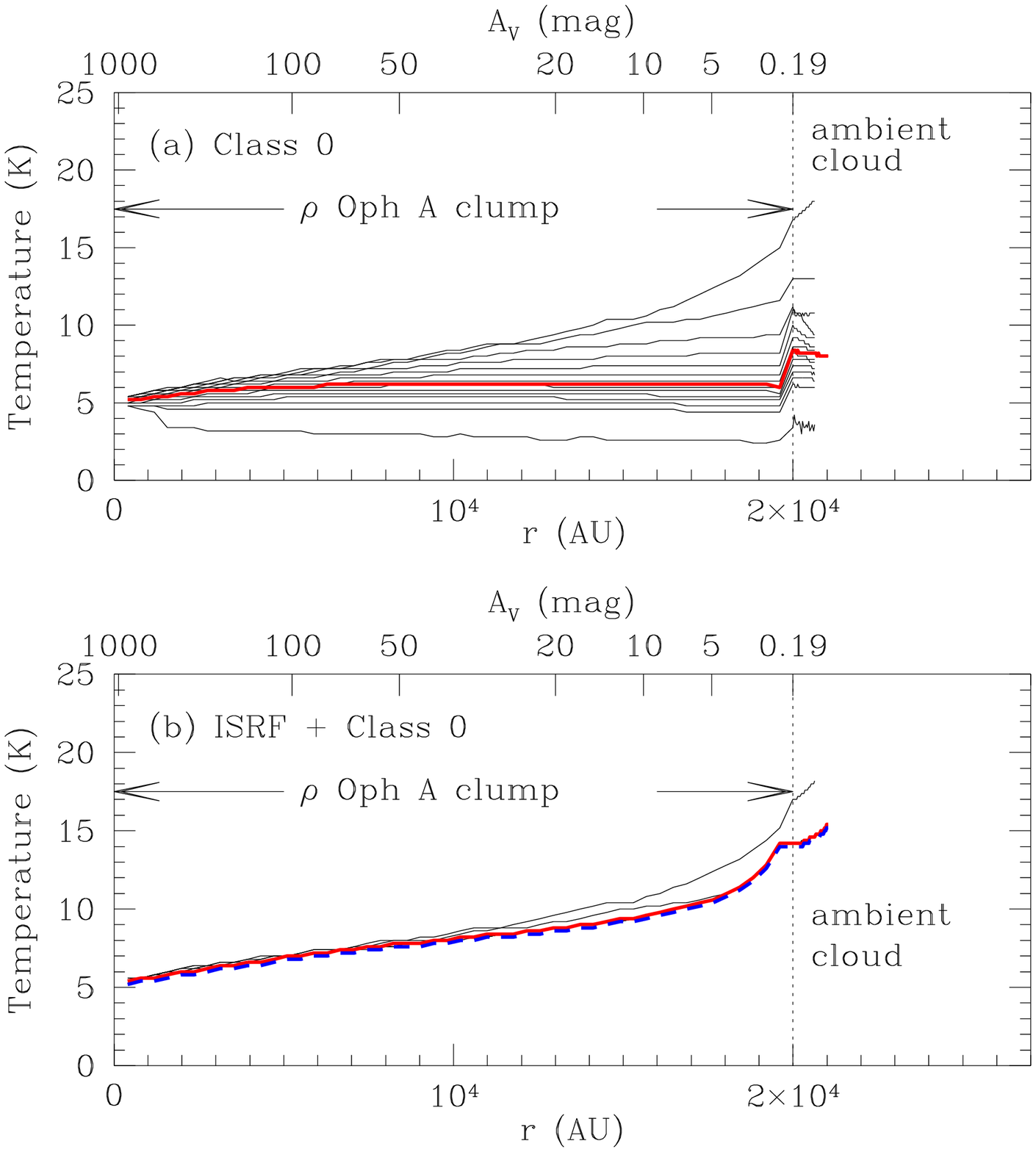}}
\caption{Same as in Fig.~\ref{fig:class0}, but for heating from a  
 Class I object at the edge of the clump. The ISRF is again dominant. }
\label{fig:classI}
\end{figure}

\subsection{$\rho $ Oph C}

The estimated temperature of this clump from previous observations is
around 10~K (e.g. Motte et al. 1998; Andr\'e et al. 1993). 
This clump is colder than the previous two, suggesting that it is
shielded from HD147879, possibly by parts of the other $\rho$ Oph clumps.
The temperature of this clump drops from 11-13~K at its edge of to
6-7~K at its centre (Fig.~\ref{fig:ophC}a). The SED of the system
peaks at around 270 $\micron$ (Fig.~\ref{fig:ophC}b), which corresponds to a temperature
of 9~K. Assuming that cores exist at visual extinctions $ A_{\rm V}>7$~mag
we find that their temperatures should be below 10~K.

\subsection{$\rho $ Oph D}
The temperature profile and the SED 
of $\rho $ Oph D are presented in Fig.~\ref{fig:ophD}.
The SED peaks at around 205~$\micron$, corresponding to a temperature
of 12~K, as assumed by Motte et al. (1998). However, most
of the clump is colder than 12~K.
Assuming that cores exist at visual extinctions$ A_{\rm V}>7$~mag
we find that their temperature should be below 10~K.

\subsection{$\rho $ Oph F}

The temperature profile and the SED 
of $\rho $ Oph F are presented in Fig.~\ref{fig:ophF}.
The SED peaks at around 160\micron, corresponding to a temperature
of 16~K. However, most of the clump is colder than 16~K, with
the region  at visual extinctions $ A_{\rm V}>7$~mag
having temperature  below 10~K.

\subsection{Dust heating by young protostars}
\label{sec:protostars}

We have so far ignored heating from nearby young protostars.
Motte et al. (1998) identify a number of Class 0, Class I and
Class II objects either near the clumps or embedded in the clumps.
The presence of these luminosity sources is expected to increase the
clump temperature. In order to quantify this increase we use
the $\rho$ Oph A clump as a study case, and
consider heating  (i) from a Class 0 object
located at the edge of the clump, (ii) from a Class I object
also located at the edge of the clump, and (iii)
from a Class I object embedded in the clump,  0.05~pc from
its centre. 
 The SEDs of the Class 0 and the Class I objects are taken
from the simulations of Stamatellos et al. (2005) (Figs.~\ref{fig:class0.sed},
 \ref{fig:classI.sed}). The bolometric luminosity 
(i.e. the integrated luminosity over the entire spectrum)
of the Class 0 source is
5~L$_{\sun}$, and that of the Class I source is 10~L$_{\sun}$. These are
typical luminosities of such objects in the region (e.g. Wilking et al. 1989).

\begin{figure}
\centerline{\includegraphics[width=7.cm]{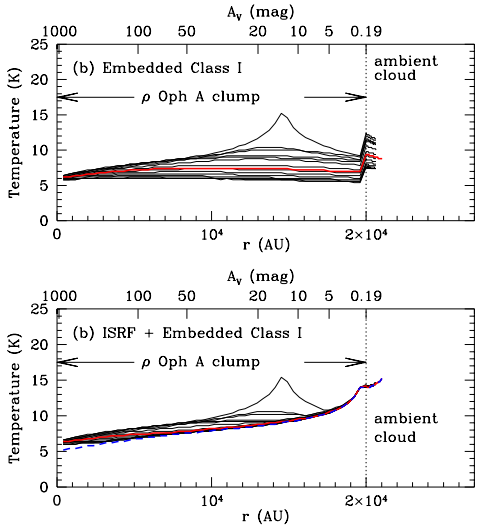}}
\caption{Same as in Fig.~\ref{fig:class0}, but for heating from a 
Class I object located inside
the clump (0.05 pc away from the centre of the clump).
The heating provided by the ISRF is still dominant apart from the region
close to the embedded protostar.}
\label{fig:classIem}
\end{figure}

The results of the radiative transfer simulations are presented in
Figs.~\ref{fig:class0}, \ref{fig:classI}, and \ref{fig:classIem}.
On the top graph of each figure we present the dust temperature taking
into account only the heating from the nearby or the embedded source. The
clump is hotter closer  to the assumed source, as expected, with the temperature
ranging from 4 to 15~K in different directions inside the clump.
On the bottom graphs of each figure
we present the dust temperature  also taking into account the heating of the ISRF, 
which has a bolometric luminosity of  5~L$_{\sun}$, i.e. comparable
with the luminosities of the nearby or embedded sources. As seen in the figures
the role of the ISRF is dominant;
the presence of  the  nearby Class 0 object does not affect the dust temperature at
all, and the presence of a nearby Class I object just heats the 
outer layers of the clump that are closer to it by only $\sim 2$~K. 
The presence of an embedded Class I object is more noticeable. 
It increases the dust temperature in the clump by  $\sim1-2$~K, and very
close to it by $\sim 5~K$. However, we note that (i) the temperature 
of most of the clump is still below 10~K, and (ii) the increase of temperature
is not greater than the increase of temperature when the radiation from HD147889
is considered (cf. Fig.~\ref{fig:ophA}b). This is because (i) the luminosity of 
the protostar is comparable to the luminosity of the ISRF, hence smaller than
the luminosity of HD147889 reaching the clump, 
and (ii) most of the protostar's radiation is emitted
at long wavelengths, where the clump is optically thin. 

We conclude that the effect of
heating due to the presence of  nearby or embedded young protostars
is not greater than the effect of the ISRF, and hence secondary to the effect of
HD147889. Thus, nearby or embedded young protostars
do not greatly affect the dust temperatures in the regions deep inside
the clumps where prestellar cores are observed, 
unless there is a large number of young protostars within
each clump, which is not the case for the $\rho$ Oph main cloud.

\subsection{The effect of a 3D clumpy structure}
\label{sec:sph}

In the radiative transfer simulations presented so far we use 
spherically symmetric models to describe the big clumps in the $\rho$ Oph main
cloud. 
In reality these clumps are more structured, as they contain regions where 
the density is either higher or lower than the assumed spherically symmetric
model. Indebetouw et al. (2006) examined the effect the cloud clumpiness in the case
of clouds that contain high-mass stars and they found that the dust temperature
is affected significantly  due to the clumpiness of the medium.

\begin{figure}
\centerline{\includegraphics[width=6.3cm,angle=-90]{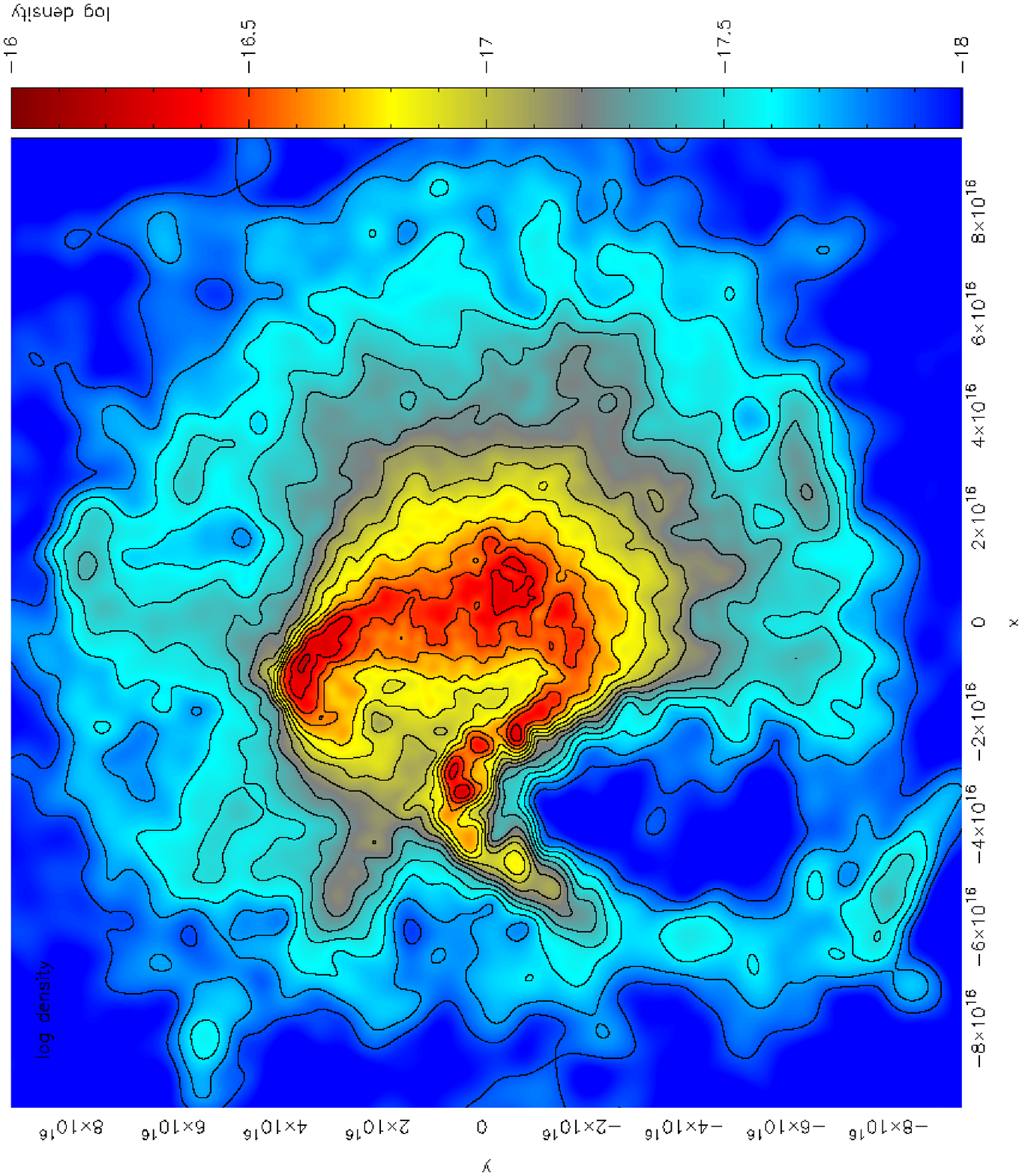}}
\caption{Logarithmic density (in ${\rm g\ cm}^{-3}$) on the $x-y$ plane 
of a simulated Oph A-like clump. The density ranges from $10^{-18}$ to 
$10^{-16}$ ${\rm g\ cm}^{-3}$. x and y are given in cm.}
\label{fig:ophA3D}
\centerline{\includegraphics[width=7.3cm]{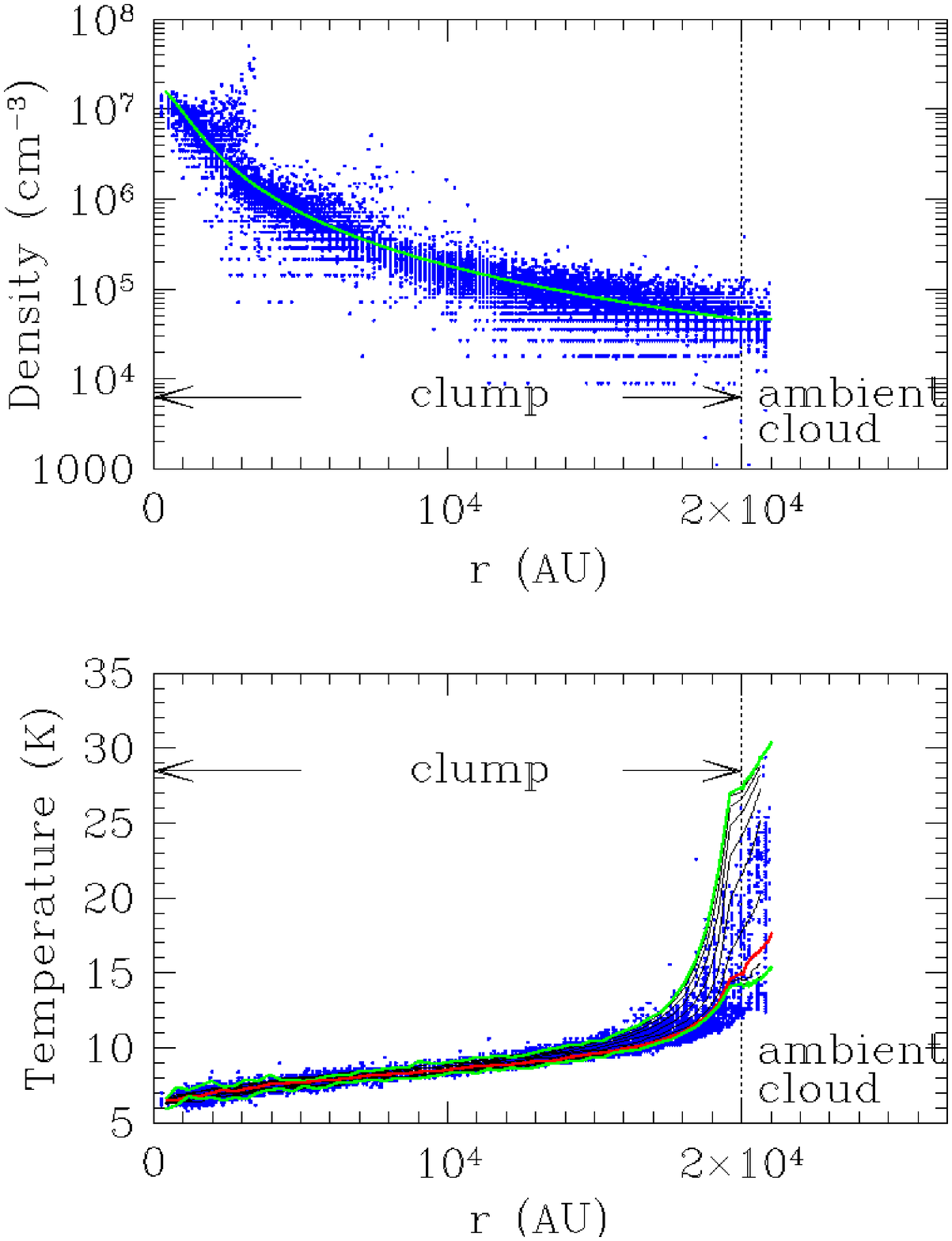}}
\caption{Density and temperature profiles (dots) of the simulated clump of 
Fig.~\ref{fig:ophA3D}. Each dot corresponds to the temperature of each radiative transfer
cell constructed from the output of the SPH simulation.
The density and temperature profiles of the spherically
symmetric model are also plotted (lines). The upper green line corresponds to the direction
towards HD147889 and the lower green line to the opposite direction.
The dust temperature in the case of 
a 3D clumpy structure is similar to the temperature calculated using the
spherical symmetric model.}
\label{fig:ophA3Ddt}
\end{figure}

To examine the effect of the 3D clumpy structure of the regions of the $\rho$ Oph 
main cloud, we generate a 3D $\rho$ Oph-A-like clump as follows. We
start off with a spherically symmetric clump with the parameters adopted for
$\rho$ Oph-A (see Table~\ref{tab:model.params}). We then impose a turbulent velocity 
field (e.g. Goodwin et al. 2004), and use {\sc DRAGON}, an SPH code, to follow the 
evolution of the cloud. Due to the effect of turbulence the cloud acquires 
a clumpy structure. The 
cloud is evolved until cores are formed (Figs.~\ref{fig:ophA3D}, \ref{fig:ophA3Ddt}).
We then perform a radiative transfer simulation on the clumpy structure using 
the method of Stamatellos \& Whitworth (2005a). As in the previous sections we
consider heating both by the ISRF and HD147889. The calculated 
 temperature profile is presented in  Fig.~\ref{fig:ophA3Ddt}. The dust
temperature is similar to the temperature calculated in the spherically symmetric case,
despite the fact that the density in some regions of the cloud is  up to an
order of magnitude different.
This is because the heating of the dense parts of the cloud is mainly due to long
wavelength radiation that propagates into the cloud without ``seeing'' the clumpy 
structure;
short wavelength radiation is  absorbed in the outer layers
of the cloud and re-emitted at longer wavelengths.

We conclude that the clumpiness of the cloud does not significantly affect 
the dust temperature at the inner, dense  regions of the cloud.

\subsection{The dust temperature of the cores in the $\rho$ Oph main cloud}
\label{newtemps2}

According to the model presented here, the dust temperatures of the cores in  
$\rho$ Oph main cloud are lower than  previously thought.  Since cores are 
observed at visual extinctions $>7$ mag, the temperatures\ of the cores in 
Oph-A  are most probably  below 11~K, and in Oph-B2, Oph-C, Oph-D and Oph-F below 
10~K. These temperatures correspond to the temperature of the clumps where the cores
are embedded, at visual extinctions $A_{\rm V}=7$ mag.

Considering the fact
that most of the cores are observed at visual extinctions 
$12-20$ mag (Johnstone et al. 2004), 
their actual temperatures may be even lower by $\sim~2$~K. 
Hence, the presence of the luminous   HD147889 in the vicinity of the
clumps, affects only the outer regions of the clumps, where 
prestellar cores have not been observed.

\section{The core mass function of $\rho$ Oph}

Motte et al. (1998) calculated the core mass function of the $\rho$ Oph main cloud by 
assuming a temperature of 20~K for Oph-A, 12~K for Oph-B, Oph-C, and
Oph-D, and  15~K for Oph-E and Oph-F.
Johnstone et al. (2000) have assumed similar temperatures to  Motte et al. (1993)
or even higher temperatures. 
These temperatures are higher than the ones predicted by our model, by $2-9$~K.
Hence, considering the fact that core masses are calculated using 
Eq.~\ref{eq:col.dens.gen},  
we suggest that both these authors have underestimated the masses of the 
prestellar cores in this region. At these  temperatures the 
Rayleigh-Jeans relation is not a good approximation and the  Planck function
must be used; hence even overestimating  temperatures by a few K leads
to underestimating masses by a factor of $2-3$. 

\begin{figure}
\centerline{
\includegraphics[width=5.8cm,angle=-90]{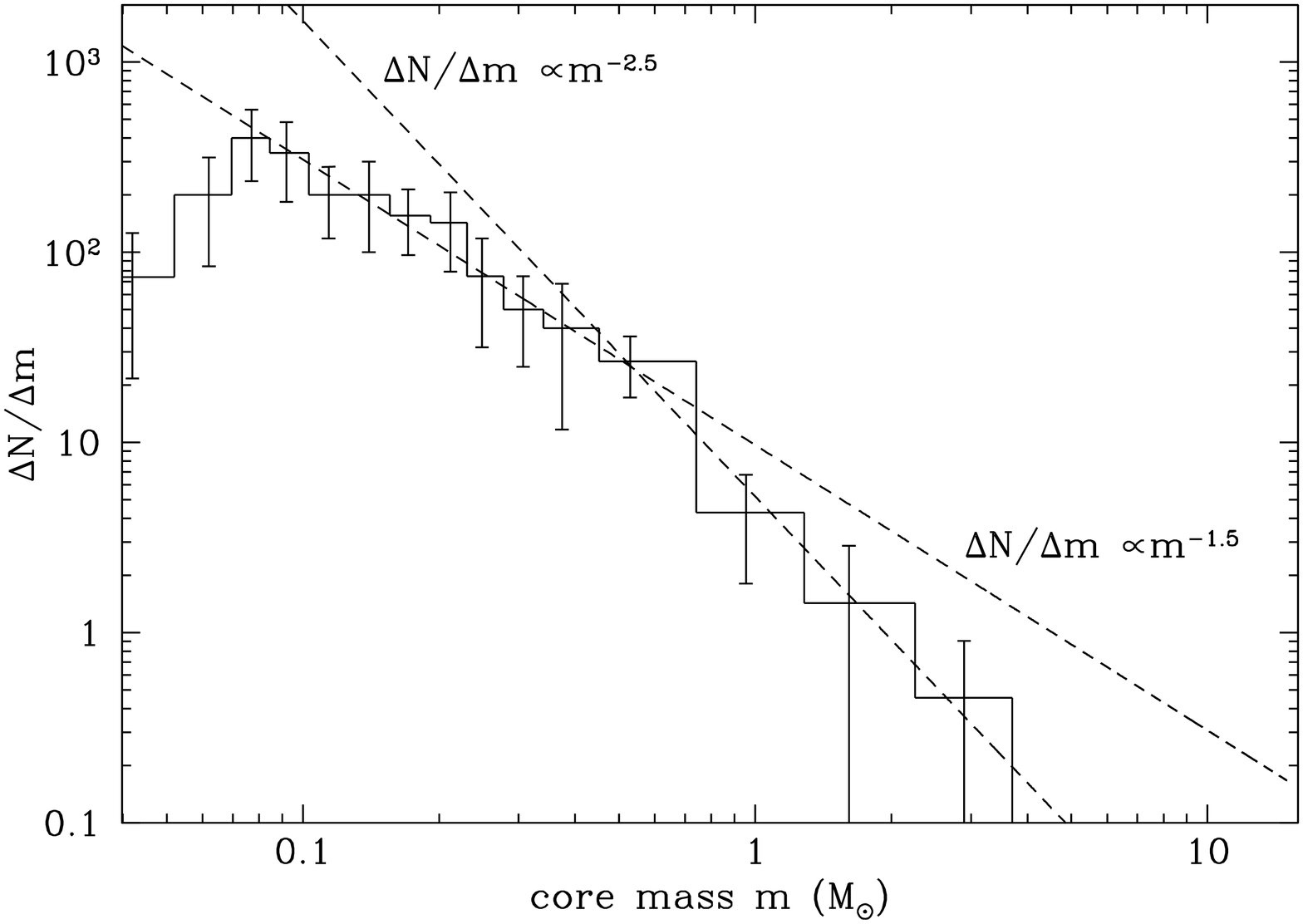}}
\caption{The core mass function of the $\rho$ Oph main cloud
based on the core masses calculated by Motte et al. (1998).
The error bars correspond to $\sqrt{N}$ counting statistics}
\label{fig:imf}
\centerline{
\includegraphics[width=5.8cm,angle=-90]{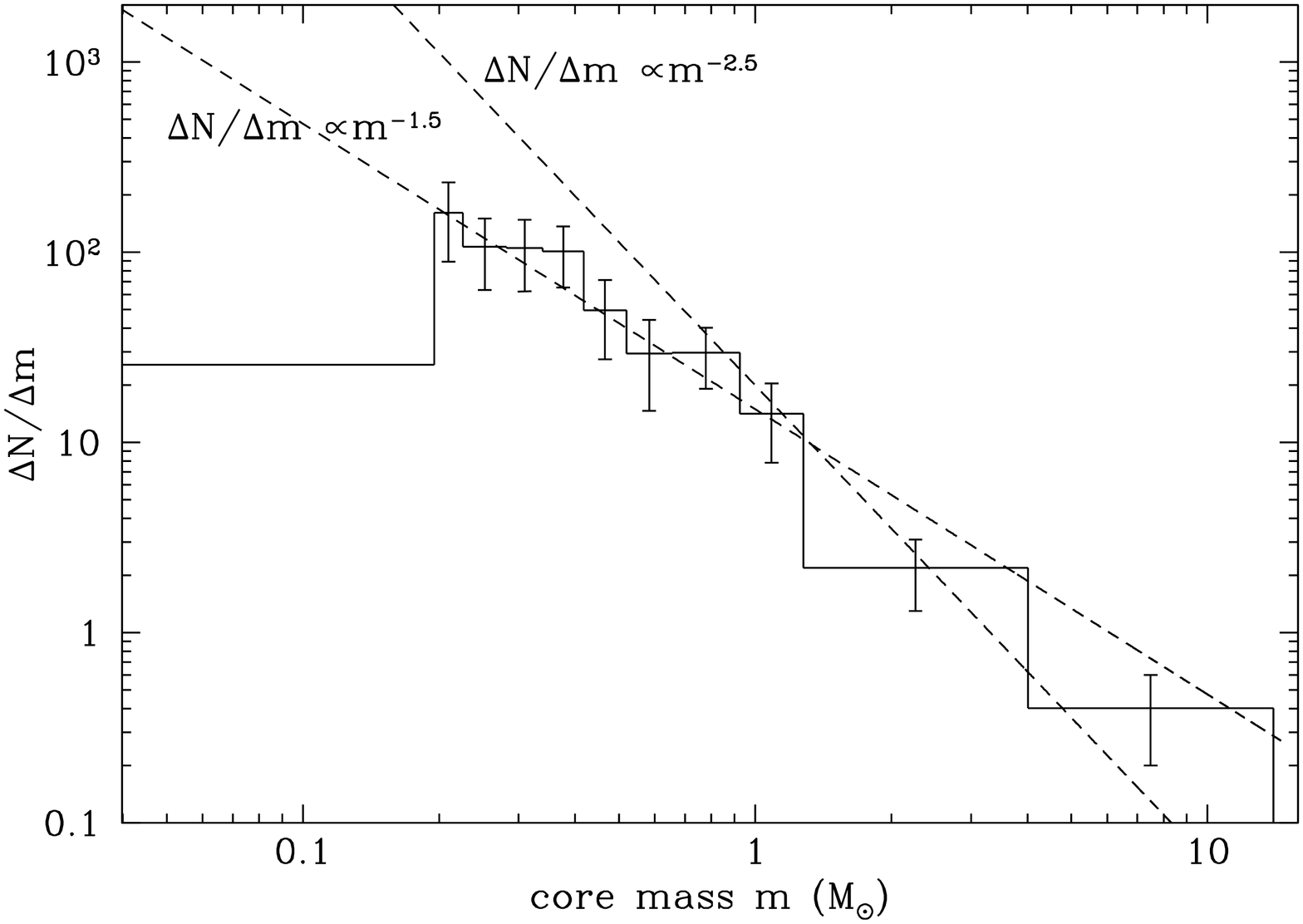}}
\caption{The CMF  of the $\rho$ Oph main cloud based on
the core masses calculated using temperatures from our model. The CMF is constructed
so that there is approximately equal number of cores per bin. The CMF has
moved to larger masses but its overall shape has not changed significantly.}
\label{fig:newimf}
\end{figure}

 We revise the core mass function of Motte et al. (1998) assuming core temperatures
derived from our model (Section~\ref{newtemps2}).  We shall assume the same dust opacity 
at 1.3~mm, i.e. $\kappa=0.005~{\rm cm}^2\ {\rm g}^{-1}$ (Preibisch et al. 1993;
Andr\'e et al. 1996). We then calculate the core the core masses from the
observed fluxes using Eq~\ref{eq:col.dens.gen}.
We note again that  temperatures we assume correspond to  cores
at visual extinctions of $\sim7$ mag. Johnstone et al. (2004) suggest that 
most of the  cores in this region are embedded at $12-21$ mag or at even higher
extinctions. Hence the temperatures of many of the cores 
are expected to be even lower (by $\sim 2$~K).  
This means that the temperatures used are upper limits to the actual temperatures,
and consequently that the computed masses are lower limits to the actual core masses.

In Fig.~\ref{fig:imf} we present the Motte et al. (1993) CMF, and in 
Fig.~\ref{fig:newimf} the revised CMF derived using core temperatures from the model 
presented in this paper. The CMF has moved to higher masses but its overall
shape has not changed significantly. The mass at which the core mass
spectrum  steepens from a slope $\alpha\sim1.5$ to a slope $\alpha\sim2.5$ is less
clear in this case but it seems that it has
moved from  $\sim0.5$~M$_{\sun}$ to  $\sim1$~M$_{\sun}$.
Contrary to the CMF in other star forming regions
(e.g. in Orion; Nutter et al. 2007)
the CMF does not show a turnover down to the completeness limit
(which is $\sim0.2$~M$_{\sun}$, using the new temperatures). 
However, the CMF may flatten at around $\sim 0.4$~M$_{\sun}$.

\section{Dust temperatures of cores in other star forming regions}

The dust temperatures of prestellar cores are  determined by the
ambient radiation field heating the core (since by definition
there are no radiation sources inside the core). 
This radiation field is determined by  environmental factors. It 
depends (i) on how deeply the core is embedded in its ambient cloud,
and (ii) on the possible presence of 
nearby radiation sources,
and their relative position/distance with respect to the core.
Hence, the external radiation field that heats each core
is different. 

Previous studies of prestellar cores and young protostars
(e.g. Evans et al. 2001; Young et al. 2003; 
Jorgensen et al. 2006) have acknowledged this fact and 
have used a scaled version of the standard ISRF that is 
either enhanced at all wavelengths or selectively at UV and FIR. 
This simple approach has a free parameter, 
the ISRF scaling factor, which is varied arbitrarily to fit the observations but
it is not connected directly to the molecular cloud in which the core is
embedded or the transport of radiation inside the cloud. It also
does not account for the fact that the radiation field incident on 
an embedded core is not isotropic.

Here, we follow the approach of Stamatellos et al. (2004) 
where the radiation field incident on the core  is
a direct result of the presence  (i) of the ambient cloud that
surrounds the core,  (ii) of nearby embedded young 
protostars, (iii) of nearby luminous stars.  
The ambient cloud of each core/clump attenuates the
ISRF, as it acts like a shield to
UV, visual and NIR interstellar radiation, absorbing 
and re-emitting it in the FIR. The nearby young protostars and stars
also enhance the radiation field incident on core, mainly at long wavelengths
($>50~\micron$) due to the thermal emission 
from the outer regions of the heated ambient cloud (Mathis et al. 1983), as
short wavelength radiation cannot penetrate deep into the cloud.

\subsection{Dust temperatures of the prestellar cores in the $\rho$ Oph molecular
cloud}

In the previous sections we discussed in detail the region of the
$\rho$ Oph main cloud, where the external heating is dominated by a nearby
B2V star. Due to the presence of this star and embedded protostars in the
$\rho$ Oph main cloud the external radiation field is stronger than the
standard ISRF by a factor of $\sim 10$ (the bolometric 
luminosity of the ISRF heating the core is $\sim 5~{\rm L}_{\sun}$
whereas the  bolometric luminosity of the star's radiation reaching and heating
each clump is $\sim 50~{\rm L}_{\sun}$). This is consistent with the
observations of Liseau et al. (1999) which suggest that 
the external radiation field incident on $\rho$  Oph  
$\sim 10-100$ times stronger than the standard ISRF.

Using a detailed model presented in the previous sections, we 
estimate that the temperatures of the cores in  the $\rho$ Oph main 
cloud are probably below 10-11~K, i.e. lower than previous estimates.
The dust temperatures of the cores in  L1689,
 another region of the Ophiuchus molecular cloud which is totally starless
(e.g. Nutter et al. 2006),  are also probably below 10-11~K.

\begin{figure}
\centerline{
\includegraphics[width=7.7cm]{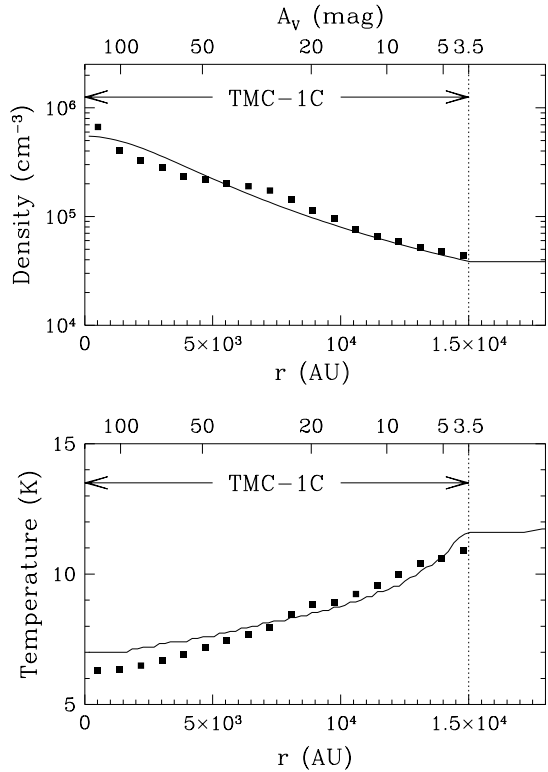}}
\caption{Assumed density profile and calculated 
temperature profile of TMC-1C (solid lines) in Taurus. 
The squares correspond to estimates based
on observations at 450~$\micron$, 850~$\micron$, and 1200~$\micron$
(Schnee et al. 2007).
 }
\label{fig:tmc}
\end{figure}

\subsection{Dust temperatures of the prestellar cores in the Taurus molecular cloud}

In Taurus there are no nearby luminous stars, hence the radiation field heating the
cores in this region is enhanced
only due to the presence of young protostars and stars in its vicinity.
This radiation field is not expected to be stronger than that in
$\rho$ Oph. Thus, the dust temperatures estimated by our model for
 the $\rho$ Oph main cloud are upper limits
to the temperatures of the cores in Taurus. 
Indeed, observations of  TMC-1C, a prestellar core in in Taurus
(Schnee \& Goodman 2005; Schnee et al. 2006),
suggest dust temperatures
that drop from 11~K at the edge of the core ($A_{\rm V}\approx10$) to 5~K
at its centre ($A_{\rm V}\approx80$), which are consistent
with heating from an ISRF that is weaker than the standard ISRF.

In Fig.~\ref{fig:tmc} we present a model for TMC-1C.
Based on the estimates from Schnee et al. (2006) 
 we assume the  density profile defined in
Eq.~\ref{plummer}, with $r_{_0}=0.02$~pc, $n_0=5.5\times 10^5$~cm$^{-3}$ and
$R_{\rm core}=15000$~AU. We further assume a virtual ambient cloud around the core
of visual extinction $A_{\rm V}=3.5$ . This virtual cloud modifies the radiation field
incident on the core. The calculated temperature (Fig.~\ref{fig:tmc}, bottom) is very
similar to the temperature estimated by Schnee et al. (2006). The calculated
temperature at the centre of the core is higher by $\sim1$~K than
the Schnee et al. (2006) estimate, indicating that the radiation field heating the core
is weaker than the standard ISRF at long wavelengths 
(or alternatively that the core centre is  denser).

We conclude that the dust temperatures of the prestellar 
cores in Taurus are probably below 10~K.

\begin{figure*}
\centerline{
\includegraphics[width=8.5cm]{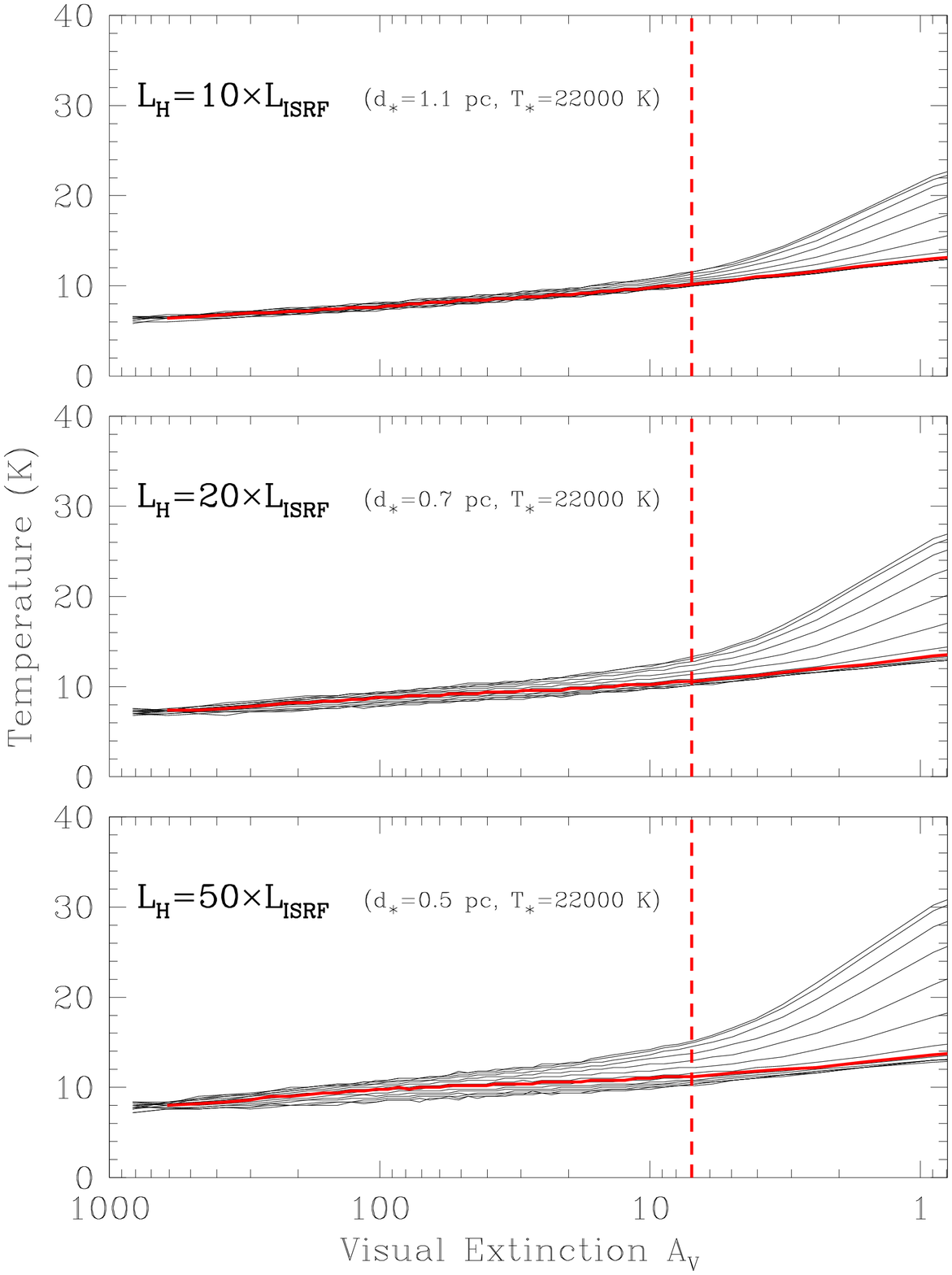}\hspace{0.3cm}
\includegraphics[width=8.5cm]{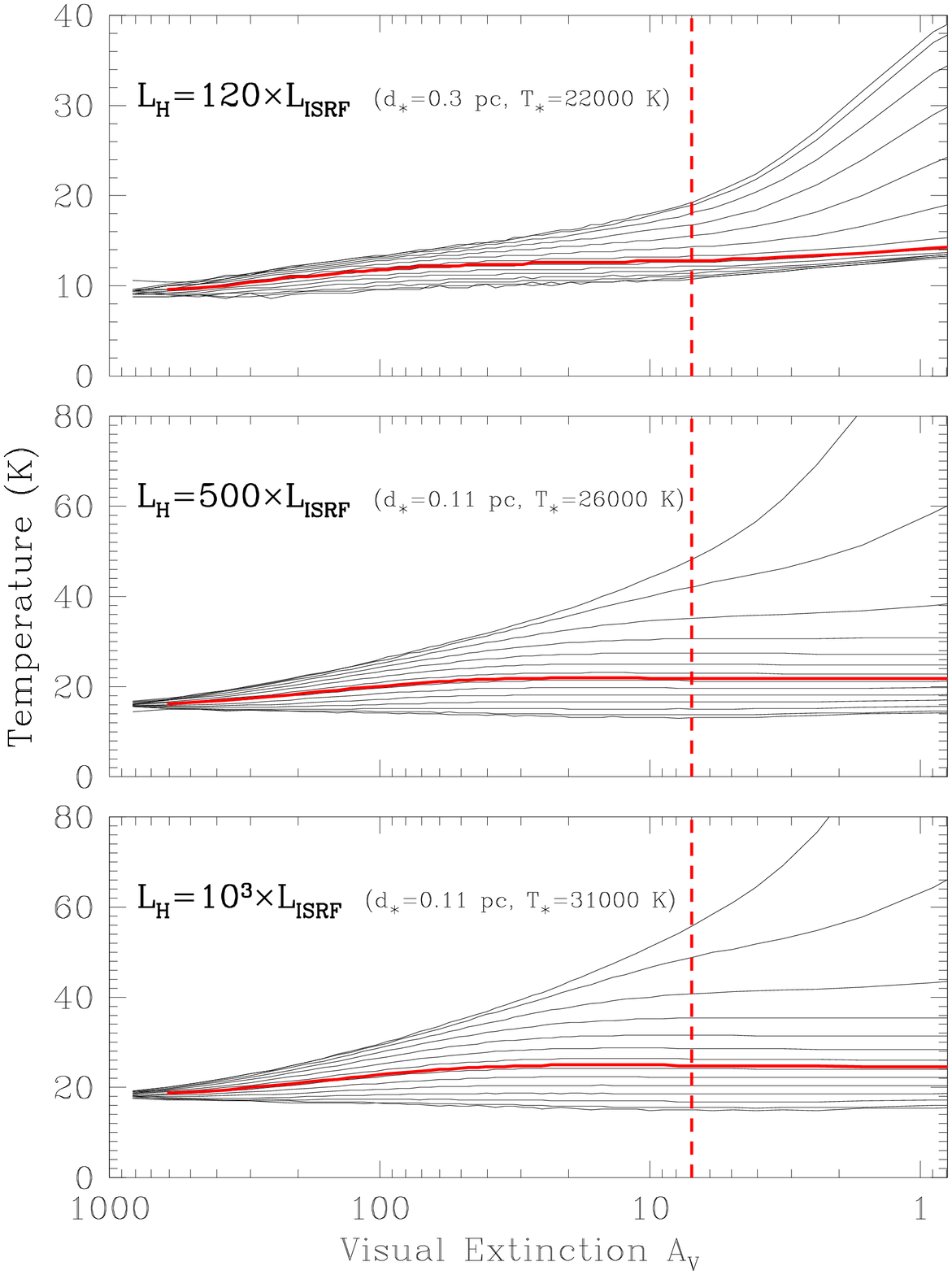}}
\caption{Dust temperature  of a $\rho$ Oph A-like clump against 
the visual extinction from the edge of the clump. The clump heating is
due to the ISRF and a B2V star having different distances from the centre
for the clump and different temperatures, so as to change the bolometric
luminosity incident on the clump (as marked on each graph). 
The temperature of most of the clump
is below 10~K unless the heating from the star is very strong 
($\stackrel{>}{_\sim}120\times L_{\rm ISRF}$). }
\label{fig:temp+}
\end{figure*}

\subsection{Dust temperatures of the prestellar cores in the Orion molecular cloud}

In Orion there are many radiation sources that contribute to the heating 
of the prestellar cores in region.
Jorgensen et al. (2006) estimates that the external radiation field in
the Orion region is up to $\sim 10^3$ stronger than the standard 
interstellar radiation field.

In order to investigate the effect of such an enhanced radiation field on the dust
temperatures in prestellar cores in  Orion, we use  $\rho$ Oph  A 
as a representative clump, and consider heating from a B2V star (similar to HD147889)
with different temperatures and at different distances from the centre of the clump. 
Hence, we examine the effect of heating from an incident radiation field having 
bolometric luminosity $L_{\rm H}$  10, 20, 50, 120, 500 and $10^3$ times
the bolometric luminosity of the standard interstellar radiation field 
(Black 1994; Andr\'e et al. 2002). 

In Fig.~\ref{fig:temp+} we present the dust temperature versus
visual extinction calculated using incident radiation fields of different strengths.
For radiation fields with bolometric luminosities up $50\times L_{\rm ISRF}$ the
dust temperature profile is similar to the ones calculated for $\rho$ Oph;
most of the dust at optical visual extinctions $>7$~mag is colder than 10~K. 
For a radiation field  $120\times L_{\rm ISRF}$, half of the clump
(i.e. the hemisphere away from the luminosity source) is colder than
12~K. For a higher external radiation,  $(500-10^3)\times L_{\rm ISRF}$,
the dust temperature in the clump varies considerably with the position in 
the clump; from 14 to 45~K in the case of  $500\times L_{\rm ISRF}$, 
and from 15 to 53~K  for the case of $10^3\times L_{\rm ISRF}$. 
Hence, the dust temperature of a core depends on the core position
in the clump, i.e. its relative position with respect to the external radiation source
and on how deeply embedded the source is in its parent cloud.  

We conclude that assuming that (i) the incident radiation field on the clumps of Orion is
 $(500-10^3)\times L_{\rm ISRF}$, and (ii) that cores exist at visual extinctions
 $>7$~mag, then typical dust temperatures of the prestellar cores in Orion
are from 20 to 30~K, which is consistent with previous 
assumptions/estimates (Launhardt et al. 1996; Mitchell et al. 2001;
Motte et al. 2001;  Johnstone et al. 2006; Nutter et al. 2007).

\section{Conclusions}

We have used a 3D geometry for the $\rho$ Oph main cloud region
to construct a radiative transfer model for this region, taking into account
external heating by  (i) the interstellar radiation field, and (ii) 
HD147889, a nearby B2V star. HD147889 dominates
the heating of the $\rho$ Oph main cloud clumps.  We estimate that
the dust temperatures at visual extinctions $>7$~mag are
below $\sim$10-11~K. These are smaller than was previously assumed.
As a result we find that the core masses calculated from mm observations
are underestimated by a factor of 2-3. This affects the core mass function 
of the  $\rho$ Oph main cloud. We computed a revised CMF for this region
using the dust temperatures calculated in this paper.
 The CMF has moved to higher masses but its 
shape has not changed significantly. The mass at which the core mass
spectrum  steepens from a slope $\alpha\sim1.5$ to a slope $\alpha\sim2.5$ is less
clear but it appears that it has
moved from  $\sim0.5$~M$_{\sun}$ to  $\sim1$~M$_{\sun}$.
This is still below the mass where this steepening occurs in Orion.
Contrary to the CMF in Orion
the CMF of the  $\rho$ Oph main cloud does not show a turnover at low masses. 
However, it may flatten at around $\sim 0.4$~M$_{\sun}$.

We have generalized our study to estimate the dust temperatures
in  prestellar cores in other star forming regions.
In the Taurus molecular cloud the ambient radiation field is weaker than that
in Ophiuchus, hence the dust temperatures of the cores in this region are similar
or smaller than the ones calculated for $\rho$ Oph. 
We estimate that the dust temperatures at visual extinctions $>7$~mag are
 below $\sim$10~K.
In Orion the ambient radiation field is estimated to be up to $10^3$ times stronger
than the standard interstellar radiation field. Using a simple model to account for this
enhanced radiation field, 
we find that the typical dust temperatures of the prestellar cores in this region
are from 20 to 30~K.

\section*{Acknowledgements}
   
We would like to thank S. Schnee for providing the data for TMC-1C, 
and P. Andr\`e for providing an improved version of the BISRF. 
We also thank J.~Kirk  and D.~Nutter
for useful discussions on prestellar cores in Taurus and Orion, and
R.~Simpson for Fig.~\ref{fig:imf}. 
We acknowledge support by PPARC grant PPA/G/O/2002/00497. 


\end{document}